\documentclass[aoas,MSNbibl,nameyear,noautosecdot,dvips]{arximspdf}
\usepackage{algorithm}
\usepackage{dcolumn}
\usepackage{graphicx}

\doi{10.1214/14-AOAS781} % kopijuoti is laisko
\volume{8}
\issue{4}
\pubyear{2014}
\firstpage{2435}
\lastpage{2460}
\docsubty{FLA}

\makeatletter
\newcolumntype{d}[1]{D{.}{.}{#1}}
\newcommand{\rrvert}{\vert}
\newcommand{\rrVert}{\Vert}
\newcommand{\llvert}{\vert}
\newcommand{\llVert}{\Vert}
\newcommand{\bzero}{\mathbf{0}}
\newcommand{\bb}{\mathbf{b}}
\newcommand{\bh}{\mathbf{h}}
\newcommand{\bR}{\mathbf{R}}
\newcommand{\bY}{\mathbf{Y}}
\newcommand{\bff}{{\mathbf f}}
\newcommand{\hf}{\hat{\bff}}
\newcommand{\bfeta}{\bolds\eta}
\newcommand{\bomega}{\bolds\omega}
\newcommand{\bxi}{\bolds\xi}
\newcommand{\hbxi}{{\hat{\bxi}}}
\newcommand{\balpha}{{\bolds{\alpha}}}
\newcommand{\btheta}{{\bolds{\theta}}}
\newcommand{\bphi}{{\bolds{\phi}}}
\makeatother

\begin{document}
\begin{frontmatter}

\title{Functional response additive model estimation with~online virtual stock markets}
\runtitle{FRAME}

\begin{aug}
\author[A]{\fnms{Yingying}~\snm{Fan}\thanksref{M1,T1}},
\author[B]{\fnms{Natasha}~\snm{Foutz}\thanksref{M2}},\\
\author[A]{\fnms{Gareth M.}~\snm{James}\thanksref{M1}}
\and
\author[D]{\fnms{Wolfgang}~\snm{Jank}\corref{}\thanksref{M3}\ead[label=e1]{wjank@usf.edu}}
\runauthor{Fan, Foutz, James and Jank}
\affiliation{University of Southern California\thanksmark{M1},
University of Virginia\thanksmark{M2}\\ and
University of South Florida\thanksmark{M3}}
\address[A]{Y. Fan\\
G. M. James\\
University of Southern California\\
Bridge Hall 401\\
Los Angeles, California 90089-0809\\
USA}
\address[B]{N. Foutz\\
McIntire School of Commerce\\
University of Virginia\\
340 Rouss and Robertson Hall\\
Charlottesville, Virginia 22904\\
USA}
\address[D]{W. Jank\\
Information Systems Decision Sciences Department\\
University of South Florida\\
4202 E. Fowler Avenue\\
BSN 3403\\
Tampa, Florida 33620\\
USA\\
\printead{e1}}
\end{aug}
\thankstext{T1}{Supported in part by NSF CAREER Award DMS-11-50318.}

% HISTORY:
\received{\smonth{8} \syear{2013}}
\revised{\smonth{7} \syear{2014}}

% ABSTRACT
%
\begin{abstract}
While functional regression models have received increasing attention recently,
most existing approaches assume both a linear relationship and a scalar
response variable.
We suggest a new method, ``Functional Response Additive Model Estimation'' (FRAME), which
extends the usual linear regression model to situations involving both
functional predictors,
$X_j(t)$, scalar predictors, $Z_k$, and functional responses, $Y(s)$.
Our approach uses a
penalized least squares optimization criterion to automatically perform
variable selection in situations involving multiple functional and
scalar predictors. In addition, our method uses an efficient coordinate
descent algorithm to fit general nonlinear additive relationships
between the predictors and response.

We develop our model for novel forecasting challenges in the
entertainment industry.
In particular, we set out to model the decay rate of demand for
Hollywood movies using the
predictive power of online virtual stock markets (VSMs). VSMs are
online communities that, in a market-like fashion, gather the
crowds' prediction about demand for a particular product. Our fully functional
model captures the pattern of pre-release VSM trading prices and provides
superior predictive accuracy of a movie's post-release demand
in comparison to traditional methods. In addition, we propose graphical tools
which give a glimpse into the causal relationship between market behavior
and box office revenue patterns, and hence provide valuable insight to
movie decision makers.
\end{abstract}

% KEYWORDS
\end{frontmatter}

%%%%%%%%%%%%%%%%%%%%%%%%%%%%%%%%%%%%%%%%%%%%%%%%%%%%%%%%%%%%%%%%%%
%%%%%%%%%%%%%%%%%%%%%%%%%%%%%%%%%%%%%%%%%%%%%%%%%%%%%%%%%%%%%%%%%%
%%%%%%%%%%%%%%%%%%%%%%%%%%%%%%%%%%%%%%%%%%%%%%%%%%%%%%%%%%%%%%%%%%
%s1 #&#
\section{\texorpdfstring{Introduction.}{Introduction}}

Functional data analysis (FDA) has become an important topic of study
in recent years, in part
because of its ability to capture patterns and shapes in a parsimonious
and automated fashion [\citet{RamsaySilverman2005}].
%Some of the areas in which FDA has been applied include
Recent methodological advances in FDA include functional principal
components analysis [\citet{james1,rice1paper}], regression with
functional responses [\citet{zeger1}] or functional predictors [\citet
{ferraty1,james5}], functional linear discriminant analysis [\citet
{james2,ferraty2}], functional clustering [\citet{james4,barjoseph1}]
or functional forecasting [\citet{zhangjankshmueli2010}].

In this paper we are interested in the regression situation involving
$p$ different functional predictors, $X_{1}(t),\ldots, X_{p}(t)$. Most
existing functional regression models assume a \textit{linear}
relationship between the response and predictors [\citet{yao2}], which
is often an overly restrictive assumption. Recently, several papers
have suggested approaches for performing nonlinear functional
regressions [\citet{james5,chen2,fan1}] of the form
%
%e1 #&#
\begin{equation}
\label{far} Y_i = \sum_{j=1}^p
f_j (X_{ij} )+ \varepsilon_i, \qquad i = 1,\ldots, n,
\end{equation}
where the $f_j$'s are general nonlinear functions of $X_{ij}(t)$ and
$Y_i$ is a centered response. Generally speaking, these approaches
operationalize estimation of equation (\ref{far}) by using functional
index models.
%Their method uses a penalized least squares criterion and is capable
%of automatically performing variable selection even for very large
%values of $p$. While the approach of Fan and James has several
%desirable properties it is only designed for data with scalar
%responses. The data that motivated our research includes not only
%functional predictors but also functional responses.
While all of these approaches provide a very flexible extension of the
linear functional model, they are designed for scalar responses only.
In this paper, we generalize this framework to functional responses.
That is, we consider both functional predictors $X_{ij}(t)$ and
functional responses $Y_i(s)$ and allow them to be related in a
nonlinear way.

%In this article we propose a new

We refer to our proposed nonlinear functional regression method as
``Functional Response Additive Model Estimation'' (FRAME), which models
both multiple functional predictors as well as functional responses.
Beyond the extension to functional responses, FRAME makes two
additional important contributions to the existing literature.
First, it uses a penalized least squares approach to efficiently fit
high-dimensional functional models while simultaneously performing
variable selection to identify the relevant predictors, an area that
has received very little attention in the functional domain. FRAME is
computationally tractable because we use a highly efficient coordinate
descent algorithm to optimize our criterion. Second, FRAME extends
beyond the standard linear regression setting to fit general nonlinear
additive models. Since the predictors, $X_{ij}(t)$, are infinite
dimensional, any functional regression model must perform some kind of
dimension reduction. FRAME achieves this goal by modeling the response
as a nonlinear function of a one-dimensional linear projection of
$X_{ij}(t)$, a functional version of the \emph{single index model}
approach. Our method uses a supervised fit to automatically project the
functional predictors into the best one-dimensional space. We believe
this is an important distinction because projecting into the
unsupervised PCA space is currently the dominant approach in functional
regressions, even though it is well known that this space need not be
optimal for predicting the response. Our nonlinear approach allows us
to model much more subtle relationships and we show that, on our data,
FRAME produces clear improvements in terms of prediction accuracy.

We develop our model for novel forecasting challenges in the motion picture
industry. Providing accurate forecasts for the success of new products is
crucial for the 500 billion dollar entertainment industries
(such as motion picture, music, TV, gaming and publishing). These
industries are % However,
%such introductions are
confronted with enormous investments, short
product life-cycles, and highly uncertain and rapidly decaying demand.
For instance, decision makers in the movie industry are keenly interested
in accurately forecasting a product's
\textit{demand pattern} %over its extremely short life-cycle
[\citet{SawhneyEliashberg1996,Bass2001}] %. For example, in the
%months
%leading up to a film's release, executives
in order to allocate, for example, weekly advertising budgets
according to the predicted \textit{rate of demand decay}, that is,
according to whether a film is expected to open big and then decay fast,
or whether it opens only moderately but decays very slowly.

\begin{figure}

\includegraphics{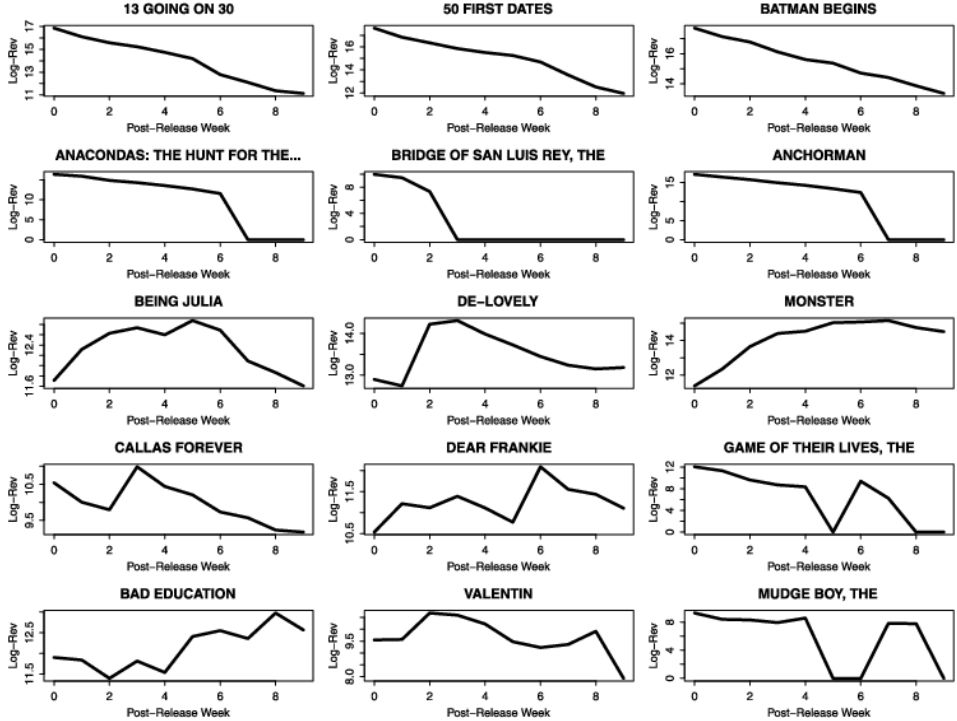}

\caption{Movie demand decay rates for a sample of movies.}\label{plotsampledecayrates}
\end{figure}

However, forecasting demand patterns is challenging since it is
highly heterogeneous across different products.
Take, for instance, the sample of movie demand patterns in
Figure~\ref{plotsampledecayrates}. Here we have plotted the log
weekly box office revenues for the first ten weeks from the release
date for a number of different movies. While revenues for some
movies (e.g., \textit{13 GOING ON 30} and \textit{50 FIRST DATES}) decay
exponentially over time, revenues for others (e.g., \textit{BEING
JULIA}) increase first before decreasing later. Even for movies with
similar demand patterns (e.g., those on the second row of Figure~\ref
{plotsampledecayrates}), the speed of decay varies greatly.

In this article we develop FRAME to forecast the demand
patterns of box office revenues using a number of functional predictors
%Our goal is to develop a flexible and powerful forecasting mechanism
%that can overcome these challenges and that can provide early and
%accurate forecasts of demand over a product's life-cycle. To
%accomplish this, we propose a forecasting model based on functional
%data analysis,
which capture various sources of information about movies, such as
consumers' word of mouth, via a novel data source, online virtual stock
markets (VSMs). %VSMs capitalize on the {\it wisdom of crowds}
In a
VSM, participants trade virtual stocks according to their
predictions of the outcome of the event represented by the stock
(e.g., the demand for an upcoming movie). As a result, VSM trading
prices can provide early and reliable demand forecasts
[\citet{Spann2003,foutzjank2010}]. %VSMs also offer monetary or
%psychological incentives (e.g. bragging rights, entertainment) to
%participants for discovering information about new products (often
%widely dispersed across media, e.g. internet, TV, trade magazines)
%and revealing this information through the embedded trading
%mechanism \citep{Dahan2007}.
%VSMs are particularly well suited for
%forecasting demand of entertainment products as they do not require
%decomposing a product into tangible attributes.
VSMs are especially
intriguing from a statistical point of view since the \textit{shape} of
the trading prices may reveal additional information, such as the
speed of information diffusion which, in turn, can proxy for
consumer sentiment and word of mouth about a new product
[\citet{foutzjank2010}]. For instance, a~last-moment price spurt may
reveal a strengthening hype for a product and may thus be essential
in forecasting its demand.

%
%f1 #&#

This paper is organized as follows. In the next section we provide
further background on virtual stock markets in general and our data
in particular. In Section~\ref{frfarsec} we present the FRAME model
and its optimization criterion. We also discuss an efficient coordinate
descent algorithm for fitting FRAME. In Section~\ref{simsec} an
extensive simulation study is used to demonstrate the superior
performance of FRAME, in comparison to a number of competitors.
Section~\ref{resultssec} discusses the results from applying FRAME to
our movie data. In that section, we also address the challenge of
interpreting the results from a model involving both functional
predictors and functional responses
%While FRAME provides high predictive accuracy, it does not easily
%allow the user to explore how and why the predictors affect the
%response.
using ``dependence plots.'' Dependence plots graphically illustrate, for
typical shapes of the predictors, the corresponding predicted response
pattern. These dependence plots allow for a glimpse into the
relationship between response and predictors and provide actionable
insight for decision makers.
%We also illustrate the insights that can be gained from our approach
%using dependence plots.
We conclude with further remarks in Section~\ref{conclusionsec}.

%describes the process of functional shape
%analysis and the insights we gain from applying shape analysis to
%the VSM trading prices as well as to the movie decay patterns.
%ection 4 discusses how the VSM and decay shapes derived in Section
%3 can be used to build our nonlinear functional forecasting model.
%We compare the accuracy of our method and

%s2 #&#
\section{\texorpdfstring{Data.}{Data}}\label{datasec}
We have two different sources of data. Our input data (i.e.,
functional predictors) come from the weekly trading histories of an
online virtual stock market for movies before their releases; our
output data (i.e.,
functional responses) pertain to the post-release weekly demand of
those movies.
We have data on a total of 262 movies. The data sources are described below.

%s2.1 #&#
\subsection{\texorpdfstring{Online virtual stock markets.}{Online virtual stock markets}}

Online virtual stock markets (VSMs) %, also known as prediction
%markets, idea futures, or betting exchanges,
operate in ways very
similar to real life stock markets except that they are not
necessarily based on real currency (i.e., participants often use
virtual currency to make trades), and that each stock corresponds to
discrete outcomes or continuous parameters of an event (rather than a
company's value). %Typical
%events often cover topics of interest to the public, such as
%economic trends (HedgeStreet), political elections (Iowa Electronic
%Markets or IEM), sporting events (TradeSports), or the Oscars
%(Hollywood Stock Exchange or HSX).
For instance, a value of \$54 for the movie stock \textit{50
FIRST DATES} is interpreted as the traders' collective belief
that the movie will accrue \$54 million in the box office during
its first four weeks of theatrical exhibition. If the movie
eventually earns \$64 million, then traders holding the
stock will liquidate (or ``cash-in'') at \$64 per share.

%For instance, a value of 54 cents for the stock ``A democratic
%candidate will win
%the Presidential election'' could be interpreted as
%%If the event has discrete outcomes (e.g. either a democratic or
%%republican candidate wins the 2008 Presidential election), then
%%trading terminates when the event occurs and the final liquidation
%%price of the stock is determined by the actual outcome of the event.
%%Therefore, the current price of a stock can be interpreted as the
%%traders' collective belief of the probability with which the event
%%will occur. For example, if the current price equals 54 cents per
%%share, then the stock can be interpreted as
%the traders' collective
%belief that the democratic candidate has a 54\% chance of winning.
%If in fact the democratic candidate wins, then traders holding the
%democratic candidate's stock will liquidate (or ``cash-in'') at \$1 per
%share; otherwise they receive \$0.

The source of our data is the \textit{Hollywood Stock Exchange} (HSX),
one of the best known online VSMs. HSX was established in 1996 and aims
at predicting a movie's revenues over its first four weeks of
theatrical exhibition. HSX has had well over 2 million active
participants worldwide and each trader is initially endowed with \$2
million virtual currency and can increase his or her net worth by
strategically selecting and trading movie stocks (such as by buying low
and selling high). Traders are further motivated by opportunities to
%exchange the accrued currency for merchandize and to
appear on the
daily \textit{Leader Board} that features the most successful traders.

\begin{figure}

\includegraphics{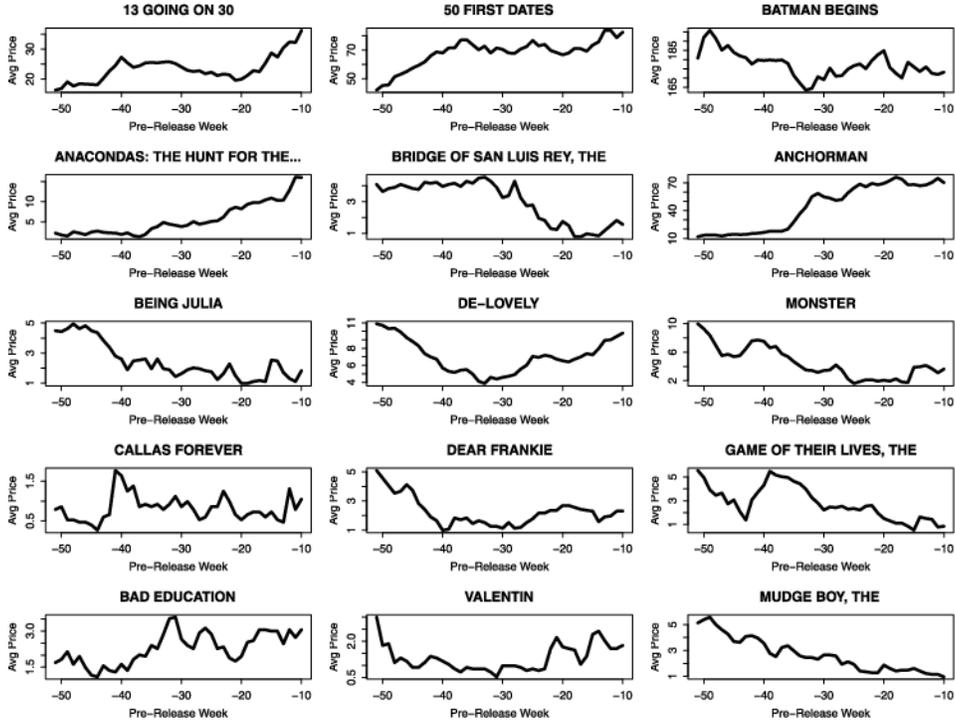}

\caption{HSX trading histories for the sample of movies from
Figure~\protect\ref{plotsampledecayrates}.} \label{plotsamplehashistories}
\end{figure}

For each movie we collect four functional predictors between 52 and 10 weeks
prior to the movie's release date. They are the following: the intra-day
average price (i.e., the average of the highest and
lowest trading prices of the day, as recorded by HSX) on each
Friday (which is the most active trading day of the week),
each Friday's number of accounts shorting the stock, number of shares
sold, and number of shares held short.
Figure~\ref{plotsamplehashistories} shows the curves for one of these
predictors, \textit{average price}, for the %sample of HSX
%trading price histories corresponding to the
movie demand patterns from Figure~\ref{plotsampledecayrates}.
%Besides trading prices, our data contains additional information about
%each stock, such as number of
%accounts trading, shares traded, and dollar volume traded.
Note that since our goal is to accomplish \textit{early forecasts}, we
only consider information between 52 and 10 weeks prior to a movie's
release (i.e., up to week $-$10 in Figure~\ref{plotsamplehashistories}). We form predictions of movie decay ten weeks prior to
release because this provides a realistic time frame for managers to
make informed decisions about marketing mix allocations and other
strategic decisions. Of course our analysis could also be performed
using data closer to the release date.

%
%f2 #&#

%
%f3 #&#
\begin{figure}

\includegraphics{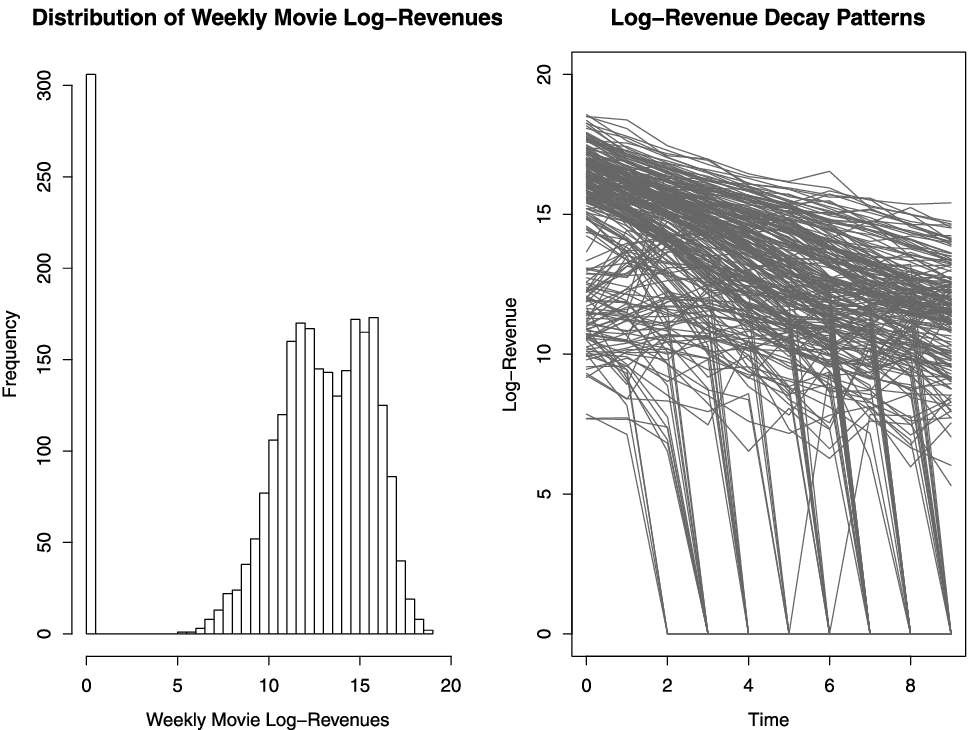}

\caption{Distribution of movies' weekly demand and demand decay patterns.
The right panel shows 10-week decay patterns (from the release
week until 9 weeks after release) for the 262 movies in our sample;
the left panel shows the distribution of the corresponding
$10\times262 ={}$2620 weekly log-revenues.}
\label{plotDist-demand-data}
\end{figure}

Our FRAME method captures differences in shapes of VSM trading
histories (such as price or volume), for example, trending up or down,
concavity vs.~convexity or last-moment spurts. The empirical results in
Section~\ref{resultssec} show that these shapes are
predictive of the demand pattern over a product's life cycle. For
example, a rapid increase in early VSM trading prices may suggest a
rapid diffusion of awareness among potential adopters and a strong
interest in a product. Thus, it can suggest a strong \textit{initial}
demand immediately after a new product's introduction to the market
place, for example, a strong opening weekend box office for a movie.
Similarly, a new product whose trading prices increase very sharply
over the pre-release period may be experiencing strong
positive word of mouth, which may lead to both a strong opening
weekend and a reduced decay rate in demand for the movie, that is,
increased longevity.

%s2.2 #&#
\subsection{\texorpdfstring{Weekly movie demand patterns.}{Weekly movie demand patterns}}

Our goal is to predict a movie's demand (i.e., its box office
revenue). Specifically, we want to predict a movie's demand not only
for a given week (e.g., at week 1 or week 5), but over its entire
theatrical life cycle of about 10 weeks (i.e., from its opening week 1
to week 10). Figure~\ref{plotDist-demand-data} shows weekly demand
for all 262 movies in our data (on the log-scale). The left panel
plots the distribution across all movies and weeks; we can see that
(log) demand is rather symmetric and appears to be bi-modal. We can
also see that a portion of the data equals zero; these correspond
to movies with zero demand, particularly in later weeks (the constant
$1$ was added to all revenues before taking the log transformation). During
weeks 1 and 2, every movie has positive revenue. In week 3, only 4
movies have zero revenue; this number increases to 67 movies by week
10. The right panel shows, for each individual movie, the rate at
which demand decays over the 10-week period. We can see that whereas
some movies decay gradually, a number have sudden drops, while others
initially increase after the release week. Our goal is to
characterize different demand decay \textit{shapes} and to use the
information from the VSM to forecast these shapes.

%In the next section, we describe a method to extract shapes from the
%daily VSM trading price histories and from the weekly movie demand
%patterns. The method is based on the ideas of functional data
%analysis and we refer to it as {\it functional shape analysis}. In
%Section 4 we then link input (prediction market) and output (movie
%demand) shapes, using different modeling alternatives including
%boosting. The full modeling process is illustrated in Figure

%s3 #&#
\section{\texorpdfstring{Functional Response Additive Model Estimation.}{Functional response additive model estimation}}\label{frfarsec}

In this section we develop our \textit{Functional Response Additive Model Estimation} (FRAME) approach for relating a functional response,
$Y_i(s)$, to a set of $p$ functional predictors, $X_{i1}(t), \ldots,
X_{ip}(t)$, and $q$ univariate predictors, $Z_{i1},\ldots, Z_{iq}$,
where $i=1,\ldots, n$.

%s3.1 #&#
\subsection{\texorpdfstring{FRAME model.}{FRAME model}}
The classical functional linear regression model is given by
%
%e2 #&#
\begin{equation}
\label{standard} Y_i(s) = \int\beta(s,t)X_{i}(t)\,dt +
\varepsilon_i(s),
\end{equation}
where $\beta(s,t)$ is a smooth two-dimensional coefficient function to
be estimated as part of the fitting process. Note we assume throughout
that the predictors and responses have been centered so that the
intercept term can be ignored. We also assume that the response curves
$Y_i(s)$ are independent, given $X_i(t)$; for work on correlated
response curves, see, for example, \citet{Chong-Zhi2009} or \citet
{Crainiceanu2011}.

The model given by (\ref{standard}) has been applied in many settings.
However, it has two obvious deficiencies for use with our data. First,
it assumes a single functional predictor, whereas our data contains
$p=4$ functional predictors and a number of univariate predictors.
Second, the integral in (\ref{standard}) is a natural analogue of the
summation term in the linear regression model. Hence, (\ref{standard})
assumes a linear relationship between the predictor and the response.
In many situations this assumption is too restrictive, so we wish to
allow for a nonlinear relationship.

%Details
%will be filled in later.}

In this paper we model the relationship between the response function
and the predictors using the following nonlinear additive model:
%
%e3 #&#
\begin{equation}
\label{generalmodel} Y_i(s) = \sum_{j=1}^p
f_j (s,X_{ij} ) + \sum_{k=1}^q
\phi _k(s,Z_{ij}) + \varepsilon_i(s),
\end{equation}
where $f_j(s,x)$ and $\phi_k(s,z)$ are general nonlinear functions to
be estimated. Model~(\ref{generalmodel}) has the advantage that it is
able to incorporate all $p+q$ predictors using a natural additive
model. It is also flexible enough to model nonlinear relationships.
However, fitting (\ref{generalmodel}) poses some significant
difficulties. First, if $p$ or $q$ are large relative to $n$, we end up
in a high-dimensional situation where many different nonlinear
functions must be estimated. We address this issue by fitting (\ref
{generalmodel}) using a penalized least squares criterion. Our
penalized approach has the effect of automatically performing variable
selection on the predictors, in a similar fashion to the lasso [\citet
{tibshirani3}] or group lasso [\citet{yuan1}] methods. Hence, we can
very effectively deal with a large number of predictors. Second, even
for a low value of $p$, estimating a completely general $f_j(s,x)$ is
infeasible because $X_{ij}(t)$ is itself an infinite-dimensional
function. Instead we model $f_j(s,x)$ using a \textit{functional single
index model}:
\[
f_j(s,X_{ij}) = g_j \biggl(\int
\beta_j(s,t)X_{ij}(t)\,dt \biggr),
\]
where $\beta_j(s,t)$ is a two-dimensional index function which
projects $X_{ij}(t)$ into a single direction and $g_j(x)$ is a
one-dimensional function representing the nonlinear impact of the
projection on $Y_i(s)$. In this way the task of estimating $f_j(s,x)$
is reduced to the simpler problem of estimating $\beta_j(s,t)$ and
$g_j(x)$. Note that our primary interest in this paper is in forming
accurate predictions for the response, $Y_i(s)$. Hence, we are
generally not concerned with identifiability of $g_j(x)$ and $\beta
_j(s,t)$, which would be more important in an inference setting.
Nevertheless, empirically we have found that $g_j(x)$ and $\beta
_j(s,t)$ can often be well estimated.

Using this functional index model (\ref{generalmodel}) reduces to
%
%e4 #&#
\begin{equation}
\label{lessgeneral} Y_i(s) = \sum_{j=1}^p
g_j \biggl(\int\beta_j(s,t)X_{ij}(t)\,dt
\biggr) + \sum_{k=1}^q
\phi_k(s,Z_{ij})+ \varepsilon_i(s).
\end{equation}
We then model $\beta_j(s,t) = \bb(s,t)^T\bfeta_j$ and
$X_{ij}(t)=\tilde\bb(t)^T\btheta_{ij}$, where $\bb(s,t)$ and
$\tilde\bb(t)$ are appropriately chosen basis functions. In
implementation, to ensure that $\beta_j(s,t)$ and $g_j(x)$ are
identifiable, we constrain $\llVert  \bfeta_j\rrVert  =1$ for all $j$. Using this
representation,
%
%e5 #&#
\begin{equation}
\label{basis-approx} \int\beta_j(s,t)X_{ij}(t)\,dt =
\btheta{}_{ij}^T \biggl[\int\tilde\bb (t)\bb(s,t)^T
\,dt \biggr] \bfeta_j = \tilde\btheta_{ij}(s)^T
\bfeta _j,
\end{equation}
where $\tilde\btheta_{ij}(s) =  [\int\bb(s,t) \tilde\bb(t)^T
\,dt ] \btheta_{ij}$. Note that $\bfeta_j$ must be estimated as
part of the fitting process, but $\tilde\btheta_{ij}(s)$ can be
assumed known for all $s$ because $\bb(s,t)$ and $\tilde\bb(t)$ are
given, so the integral can be directly computed. In addition, $\btheta
_{ij}$ can be easily computed since $X_{ij}(t)$ is directly observed.

Using this basis representation, (\ref{lessgeneral}) becomes
%
%e6 #&#
\begin{equation}
\label{framemodel} Y_i(s) = \sum_{j=1}^p
g_j \bigl(\tilde\btheta_{ij}(s)^T
\bfeta_j \bigr) +\sum_{k=1}^q
\phi_k(s,Z_{ik})+ \varepsilon_i(s).
\end{equation}
In practice, the response function, $Y_i(s)$, will generally be
observed at a finite set of time points, $s_{i1}, \ldots, s_{in_i}$.
For example, for the box office data the revenues are observed at each
of the first ten weeks. In this situation (\ref{framemodel}) can be
represented as
%
%e7 #&#
%e8 #&#
\begin{eqnarray}\label{framediscrete}
Y_{il}= \sum_{j=1}^p
g_j \bigl( \tilde\btheta{}_{ijl}^T
\bfeta_j \bigr) + \sum_{k=1}^q
\phi_k(s_l,Z_{ik})+ \varepsilon_{il},
\nonumber\\[-9pt]\\[-9pt]
\eqntext{i=1,\ldots, n, l=1,\ldots, n_i,}
\end{eqnarray}
where $Y_{il}=Y_i(s_{il})$, $\tilde\btheta_{ijl}=\tilde\btheta
_{ij}(s_l)$ and $\varepsilon_{il}$ are assumed to be independent for all
$i$ and $l$ [conditional on $X_{ij}(t)$ and $Z_{ik}$].

%s3.2 #&#
\subsection{\texorpdfstring{FRAME optimization criterion.}{FRAME optimization criterion}}
Fitting FRAME requires estimating the unobserved parameters,
$g_{j}(x),\bfeta_{j}$ and $\phi_k(s,z)$, which we achieve using a
supervised least squares penalization approach. In particular, the
FRAME fit is produced by minimizing the following criterion over a grid
of possible values for the tuning parameter $\lambda\ge0$:
%
%
%e9 #&#
\begin{eqnarray}\label{funcfar}
&& \frac{1}2\sum_{i=1}^n
\int \Biggl\{Y_i(s) - \sum_{j=1}^pg_j
\bigl( \tilde\btheta_{ij}(s)^T\bfeta_j \bigr)-
\sum_{k=1}^q\phi _k(s,Z_{ik})
\Biggr\}^2\,ds
\nonumber\\[-8pt]\\[-8pt]\nonumber
&&\qquad {} +\lambda \Biggl( \sum_{j=1}^p
\rho \bigl(\llVert f_j\rrVert \bigr) + \sum
_{k=1}^q \rho \bigl(\llVert \phi_k
\rrVert \bigr) \Biggr),
\end{eqnarray}
where $\llVert  f_j\rrVert  ^2=\sum_{i=1}^n \int f_j(s,X_{ij})^2\,ds$ with
$f_j(s,X_{ij}) = g_j ( \tilde\btheta_{ij}(s)^T\bfeta_j )$,
$\llVert  \phi_k\rrVert  ^2= \sum_{i=1}^n \int\phi_k(s,Z_{ik})^2\,ds$ and $\rho
(\cdot)$ is a penalty function.

The first term in (\ref{funcfar}) corresponds to the squared error
between $Y_i(s)$ and the FRAME prediction, integrated over $s$, and
ensures an accurate fit to the data. The second term places a penalty
on the $\ell_2$ norms of the $f_{j}(x)$'s and $\phi_k(s,z)$'s.
Note that penalizing the squared $\ell_2$ norms, $\llVert  f_j\rrVert  ^2$ and $\llVert
\phi_k\rrVert  ^2$, would be analogous to performing ridge regression.
However, we are penalizing the square root of this quantity, which has
the effect of shrinking some of the functions exactly to zero and hence
performing variable selection in a similar fashion to the group lasso
[\citet{yuan1,simon1}].
%However, we are penalizing the square root of this quantity which is
%analogous to the $\ell_1$ norm used in the lasso. This penalty has the
%effect of shrinking some of the functions exactly to zero and hence
%performing variable selection in a similar fashion to the group lasso

For a response sampled at a finite set of evenly spaced time points,
$s_1,s_2,\ldots, s_L$, we approximate (\ref{funcfar}) by
%
%
%e10 #&#
\begin{eqnarray}\label{funcfarfinite}
&& \frac{1}{2L}\sum_{i=1}^n
\sum_{l=1}^{L} \Biggl\{Y_{il} -
\sum_{j=1}^p g_j \bigl(\tilde
\btheta{}_{ijl}^T\bfeta_j \bigr) - \sum
_{k=1}^q \phi(s_l,Z_{ik})
\Biggr\}^2
\nonumber\\[-8pt]\\[-8pt]\nonumber
&&\qquad{} +\lambda \Biggl(\sum_{j=1}^p
\rho \bigl(\llVert \bff_j\rrVert \bigr)+\sum
_{k=1}^q \rho \bigl(\llVert \bphi_k
\rrVert \bigr) \Biggr),
\end{eqnarray}
where $Y_{il}=Y_i(s_l)$,
$\tilde\btheta_{ijl}=\tilde\btheta_{ij}(s_l)$,
$\llVert  \bff_j\rrVert  ^2 = \sum_{i=1}^n\sum_{l=1}^{L}g_j (\tilde\btheta{}_{ijl}^T\bfeta_j  )^2$ and $\llVert  \bphi_k\rrVert  ^2= \sum_{i=1}^n\sum_{l=1}^{L}\phi(s_l,Z_{ik})^2 $. % with $f_{ijl}=g_j\left(\tilde
Note that in using (\ref{funcfarfinite}) we are implicitly assuming
that the response has been sampled at a dense enough set of points that
the integral is well approximated by the summation term. This
approximation worked well for our data, but for sparsely sampled
responses one would need to first fit a smooth approximation of the
response and sample the fitted curve over a dense set of time points.

We further assume that $g_j(x)$ and $\phi_k(s,z)$ can, respectively, be well
approximated by basis functions $\bh(x)$ and $\bomega(s,z)$ such that
$g_j(x) \approx
\bh(x)^T \bxi_j$ and $\phi_k(s,z)\approx\bomega(s,z)^T\balpha_k$.
At each response time point $s_l$, let $\bh_{ijl}= \bh ( \tilde
\btheta{}_{ijl}^T\bfeta_j )$ and $\bomega_{ikl}=\bomega
(s_l,z_{ik})$ with $\tilde\btheta_{ijl}$ defined in (\ref
{funcfarfinite}). Then using this basis representation, (\ref
{funcfarfinite}) can be expressed as
%
%e11 #&#
\begin{eqnarray}\label{nonlinfar2}
&& \frac{1}{2L}\sum_{i=1}^n
\sum_{l=1}^{L} \Biggl\{Y_{il} -
\sum_{j=1}^p \bh_{ijl}^T
\bxi_j - \sum_{k=1}^q
\bomega_{ikl}^T\balpha _k \Biggr\}^2
\nonumber\\[-8pt]\\[-8pt]\nonumber
&&\qquad {}
+\lambda \Biggl( \sum_{j=1}^p \rho \Bigl(
\sqrt{\bxi_j^TH_j^TH_j
\bxi_j} \Bigr)+\sum_{k=1}^q
\rho \Bigl(\sqrt{\balpha_k^T
\Omega_k^T\Omega_k\balpha _k}
\Bigr) \Biggr),
\end{eqnarray}
where $H_j$ is a matrix with rows $\bh_{1j1}, \bh_{1j2}, \ldots, \bh
_{1jL}, \bh_{2j1},\ldots, \bh_{njL}$ and $\Omega_k$ is defined
similarly using $\bomega_{ikl}$. The FRAME fit is then produced by
minimizing (\ref{nonlinfar2}) over $\bfeta_j, \bxi_j$ and $\balpha_k$.

%$\bh_{ijl}= \bh\left( \tilde\btheta_{ijl}^T\bfeta_j\right)$, $
%matrices with $\sum_i n_i$ rows where the $il$th row equals $

%s3.3 #&#
\subsection{\texorpdfstring{FRAME optimization algorithm.}{FRAME optimization algorithm}}

For a given value of $\lambda$, we break the problem of minimizing
(\ref{nonlinfar2}) into two iterative steps, where we first estimate
$\bxi_j$ and $\balpha_k$ given $\bfeta_j$, and second estimate
$\bfeta_j$ given $\bxi_j$ and $\balpha_k$. One advantage of this
approach is that the minimization of (\ref{nonlinfar2}) in the first
step can be achieved using an efficient coordinate descent algorithm
which we summarize in Algorithm~\ref{alglin}.

\begin{algorithm}[t]
\caption{Step 1 of FRAME algorithm}\label{alglin}\label{alg1}
\begin{longlist}
\item[0.] Initialize $S^H_j= (H_j^T H_j )^{-1} H_j^T$ and
$S^\Omega_k= (\Omega_k^T\Omega_k )^{-1} \Omega_k^T$ for
$j=1,\ldots, p$ and $k=1,\ldots, q$, where the matrices $H_j$ and
$\Omega_k$ are defined in (\ref{nonlinfar2}).
\end{longlist}
For each~$j\in\{1,\ldots,p\}$ and $k\in\{1,\ldots,q\}$:
\begin{longlist}[6.]
\item[1.] Fix all $\hat{\bxi}_{j'}$ for $j'\ne j$. Compute the
residual vector $\bR_j = \bY-
\sum_{j'\ne j} H_{j'}\hat{\bxi}_{j'}- \sum_{k=1}^q\Omega_k\hat
\balpha_k$.
%estimate for $\bff_j$.
%and $\phi_j = \rho'(\left\Vert \hf_j\right\Vert )$
% where $\hf_j$ represents the most recent estimate for $\bff_j$.
%
\item[2.] Let $\hat{\bxi}_j = c_j S^H_j\bR_j$ where $c_j =
 (1-\lambda/\llVert  H_jS^H_j\bR_j\rrVert   )_+$ is a
shrinkage parameter.
\item[3.] Center $\hf_j \leftarrow\hf_j - \operatorname{mean}(\hf_j)$.
\item[4.] Fix all $\hat\balpha_{k'}$ for $k'\ne k$. Compute the
residual vector $\bR_k = \bY-
\sum_{j=1}^p H_{j}\hat{\bxi}_{j}- \sum_{k'\ne k}\Omega_{k'}\hat
\balpha_{k'}$.
%estimate for $\bff_j$.
%and $\phi_j = \rho'(\left\Vert \hf_j\right\Vert )$
% where $\hf_j$ represents the most recent estimate for $\bff_j$.
%
\item[5.] Let $\hat\balpha_k = c_k S^\Omega_k\bR_k$ where $c_k =
 (1-\lambda/\llVert  \Omega_kS^\Omega_k\bR_k\rrVert   )_+$ is a
shrinkage parameter.
\item[6.] Center $\hat\bphi_k \leftarrow\hat\bphi_k - \operatorname
{mean}(\hat\bphi_k)$.
\end{longlist}
Repeat 1 through 6 and iterate until
convergence.
\end{algorithm}

Our approach has the same general form as similar algorithms used in
other settings. In particular, arguments similar to those in \citet
{ravikumar1} and \citet{fan1} prove that Algorithm~\ref{alglin} will
minimize a penalized criterion of the form given by (\ref{nonlinfar2})
provided $\rho(t)=t$. We discuss the extension to a general penalty
function in the \hyperref[app]{Appendix}. Note that the $S_j^H$ and $S_k^\Omega$
matrices defined in Algorithm~\ref{alglin} only need to be computed
once so the calculations in 1 through 6 of Algorithm~\ref{alglin} can
all be performed efficiently.

In the second step we estimate $\bfeta_j$, given current estimates for
the $\bxi_j$'s and $\balpha_k$'s, by minimizing the sum of squares term
%
%e12 #&#
\begin{equation}
\label{etaopt} \sum_{i=1}^n \sum
_{l=1}^{L} \Biggl\{Y_{ij} - \sum
_{j=1}^p \bh \bigl( \tilde\btheta{}_{ijl}^T
\bfeta_j \bigr)^T\bxi_j- \sum
_{k=1}^q \bomega_{ikl}^T
\balpha_k \Biggr\}^2
\end{equation}
over $\bfeta_j$. Note that we do not include the penalty when
estimating $\bfeta_j$ because the $\bfeta_j$'s are providing a
direction in which to project $X_{ij}(t)$ and are thus constrained to
be norm one. Hence, applying a shrinkage term would be inappropriate.
Minimization of (\ref{etaopt}) can be approximately achieved using a
first order Taylor series approximation of $g_j(x)$. We provide the
details on this minimization
%and on computing initial values for the $\bfeta_j$'s
in the \hyperref[app]{Appendix}.

Formally, the FRAME algorithm is summarized in Algorithm~\ref{alg2}.
\begin{algorithm}[t]
\caption{FRAME algorithm}\label{algnonlin}\label{alg2}
\begin{longlist}[0.]
\item[0.] Choose initial values for $\hat{\bfeta}_j$ for $j\in\{
1,\ldots,p\}$.
%For each~$j\in\{1,...,p\}$,
%
\item[1.] Compute $\bh_{ijl}$ using the current estimates for $\bfeta
_j$. Estimate $\bxi_j$ and $\balpha_k$ using Algorithm~\ref{alg1}.
\item[2.] Conditional on the $\bxi_j$'s and $\balpha_k$'s from step~1, estimate the $\bfeta_j$'s by minimizing (\ref{etaopt}).

\item[3.] Repeat steps 1~and~2 and iterate until convergence.
\end{longlist}
\end{algorithm}
%

%s3.4 #&#
\subsection{\texorpdfstring{Tuning parameters.}{Tuning parameters}}

Fitting FRAME requires selecting the regularization parameter $\lambda
$ and the basis functions $\tilde\bb(t)$, $\bb(s,t)$, $\bh(x)$ and
$\bomega(s,t)$ defined in (\ref{basis-approx}) and (\ref
{nonlinfar2}). For our simulations and the HSX data we used cubic
splines to model $\bh(x), \tilde\bb(t)$ and $\bb(s,t)$, and a
simple linear representation for $\bomega(s,z)$ so $\phi
_k(s,z_k)=z_k\alpha_k$. We selected the dimensions of these bases
simultaneously using $10$-fold cross-validation (CV) based on
prediction error. More specifically, we chose a grid of values for the
dimension of each basis and randomly partitioned the original sample
into 10 subsamples of equal size. For each $k=1,\ldots, 10$, we used 9
subsamples to fit the model with dimensions of these bases fixed at a
given combination of the grid values, and used the remaining subsample
to calculate the prediction error. The cross-validated prediction error
is then calculated as the average prediction error over the 10
validation subsamples. Thus, for every combination of basis dimensions,
we obtained one cross-validated prediction error. %Out of the 10
%subsamples, a single subsample is used as the validation data to
%calculate the prediction error, and the remaining 9 subsamples are
%used as the training data to fit the model. Then for each combination
%of these basis dimensions, we calculated the 10-fold cross validated
%prediction error.
The final selected dimensions for these basis functions are the ones
which minimize the 10-fold cross-validated prediction error. Since the
FRAME algorithm is very efficient, this approach worked well on our data.

To compute $\lambda$, one could potentially add a grid of values for
$\lambda$ to the above \mbox{10-}fold CV,
fit FRAME over all possible combinations of the tuning parameter
values, and select the ``best'' value. However, a more efficient
approach is to compute initial estimates for $\bfeta_j$, minimize
(\ref{nonlinfar2}) over $\bxi_j$ and $\balpha_k$ for each possible
value of $\lambda$, choose the $\bxi_j$'s and $\balpha_k$'s
corresponding to the value of $\lambda$ with the lowest $10$-fold CV,
estimate the $\bfeta_j$'s for only this one set of parameters, and
iterate. This approach means that, for each iteration, the minimization
of (\ref{etaopt}) only needs to be performed for a single value of
$\lambda$. We found this approach worked well for choosing the tuning
parameters in both our simulated and real data analyses.
%The choice of $\lambda$ can be made using a variety of methods. We use
%cross-validation for our simulated and real data analyses.

%for each possible combination of basis dimensions, we found that was

%s4 #&#
\section{\texorpdfstring{Simulations.}{Simulations}}\label{simsec}

In this section we conduct a simulation study to compare the
performance of FRAME to several alternative functional approaches. We
first generated $p=6$ functional predictors using $X_{ij}(t) = \mathbf
{F}(t)\btheta_{ij} + \varepsilon_{ij}(t)$, where $\mathbf{F}(t)$ was a
$3$-dimensional Fourier basis, $\btheta_{ij}$ was simulated from a
$N({\mathbf0},\mathbf{I}_3)$ distribution, and the $\varepsilon_{ij}(t)$'s were
independent over $i,j$ and $t$ with a $N(0,0.1^2)$ distribution. Each
predictor was sampled at $150$ equally spaced time points over the
interval $t \in[0,1]$. In addition, $q=8$ scalar predictors, $Z_{ik}$,
were simulated from a standard normal distribution. Next, we generated
$\beta_j(s,t) = \beta_{j1}(s) + \beta_{j2}(t) + 0.1\beta
_{j1}(s)\beta_{j2}(t)$, where $\beta_{j1}(s)= \bb(s)^T\bfeta_{j1},
\beta_{j2}(t)= \bb(t)^T\bfeta_{j2}$, $\bb(\cdot)$ was a
$5$-dimensional cubic spline basis, and $\bfeta_{j1}$ and $\bfeta
_{j2}$ were independent $N({\mathbf0},\mathbf{I}_5)$ vectors.

The responses were generated from the model
%
%e13 #&#
\begin{eqnarray}\label{simresponse}
Y_i(s_\ell) = \sum
_{j=1}^pg_j \biggl(\int
\beta_j(s_\ell,t)X_{ij}(t)\,dt \biggr) + \sum
_{k=1}^q \gamma_k
Z_{ik} + \varepsilon _i(s_\ell),\nonumber\\[-10pt]\\[-10pt]
\eqntext{i=1,\ldots, n,}
\end{eqnarray}
where $\varepsilon_i(s_\ell)\sim N(0,0.1^2)$ and $Y_i(s_\ell)$ was
sampled at $20$ equally spaced time points $s_1, \ldots, s_L$ over the
interval $s\in[0,1]$. We set $g_1(x) = \sin(x)$, $g_2(x) = \cos(x)$
and $g_j(x)=0$ for $j=3,\ldots, 6$. Thus, only the first two
functional predictors were signal variables, with the remainder
representing noise. Similarly, we set $\gamma_1=1$ and $\gamma_k=0$
for $k=2,\ldots, 8$ so the last seven scalar predictors were noise
variables. All training data sets were generated using $n=200$ observations.

We compared \textit{FRAME} to six possible competitors. The simplest, \textit{Mean}, ignored the predictors and used the average of the training
response, at each time point $s$, to predict the responses on the test
data. This method serves as a benchmark to illustrate the improvement
in prediction accuracy that can be achieved using the predictors. The
next method was the \textit{Classical Functional Linear Regression} model
given by (\ref{standard}). We fit (\ref{standard}) by computing the
first $G$ functional principal components (FPC) for the response
function, and also the first $K$ FPCs for each predictor function. We
then used the $8$ scalar predictors and the $6K$ FPC scores from the
$6$ functional predictors to fit separate linear regressions to each of
the first $G$ FPC scores on the response. To form a final prediction
for the response function, we multiplied the estimated FPC scores by
the first $G$ principal component functions. The value of $G$, between
$1$ and $4$, and $K$, between $1$ and $3$, were both chosen using
$10$-fold cross-validation. The classical functional approach does not
automatically perform variable selection, so we also fit a variant
(\textit{PCA-L}). The only difference between \textit{Classical} and
\textit{PCA-L} is that the latter method used the group Lasso to compute the
linear regressions between the response and predictor principal
component scores and hence selected a subset of the predictors.

The fourth method, \textit{PCA-NL}, was identical to \textit{PCA-L} except
that a nonlinear generalized additive model (GAM) was used to regress
the response principal component scores on the predictor scores.
Standard GAM does not automatically perform variable selection, so we
fit \textit{PCA-NL} using a variant of SPAM [\citet{ravikumar1}], which
implements a penalized nonlinear additive model procedure and hence
selects a subset of the predictors. We used the Lasso penalty function
with the tuning parameter, $\lambda$, chosen over a grid of $20$
values via $10$-fold CV. Similarly, the dimension of the nonlinear
functions used in SPAM were chosen, between $4$ and $6$, using
$10$-fold CV.

The next method, \textit{Last Observation}, took as inputs $Z_{i1},\ldots, Z_{i8}$ plus the last observed values of $X_{ij}(t)$, that is,
$X_{i1}(t_{150}), \ldots, X_{i6}(t_{150})$. We then used the resulting
$14$ scalar predictors to estimate separate GAM regressions for the
response at each observed point, $Y(s_{1}),\ldots,Y(s_{20})$, a total
of $20$ different regressions. As with \textit{PCA-NL}, we used a variant
of SPAM to perform variable selection. While using only the last
observed time point may appear to be a naive approach, these methods
are common in situations like the HSX data, where it is often assumed
that all the information is captured at the latest time point. Hence,
we implemented this approach to illustrate the potential advantage from
incorporating the entire functional predictor.

The final comparison method, \textit{FPCA-FAR}, combined the FPCA approach
with the FAR method proposed in \citet{fan1}. FAR does not directly
correspond to our setting because it is designed for problems involving
functional predictors but only a scalar response. FPCA-FAR addresses
this limitation by producing $G$ separate FAR fits, one for each of the
first $G$ FPC scores. The FAR method has similar tuning parameters to
SPAM, which were again chosen using $10$-fold CV.

In fitting FRAME we set $\beta_j(s,t)=\beta_{j1}(s)+\beta_{j2}(t)$,
where $\beta_{j1}(s), \beta_{j2}(t)$ and $g_j(x)$ were approximated
using cubic splines. The dimension of the basis for both $\beta
_{j2}(t)$ and $\tilde\beta(t)$ was selected as the value among
$4,5,6$, which gave the lowest prediction error to $X_{ij}(t)$ on the
held-out time points. In particular, for each possible dimension we
held out every $5$th observed time point for each $X_{ij}(t)$, produced
a least squares fit using the remaining observations, and then
calculated the squared error between the observed and predicted values
of $X_{ij}(t)$ at the held-out time points. The value of $\lambda$ and
the dimensions of $\beta_{j1}(s)$ and $g_j(x)$ were all chosen using
$10$-fold CV in a similar fashion to the other comparison methods. We
set $\rho$ equal to the identity function, which corresponds to a
group lasso type penalty function.

In order to match a real-life setting, we deliberately generated the
data from a model that does not match the FRAME fit. In particular, the
true $\beta_j(s,t)$ function included an interaction term, while the
FRAME estimate was restricted to be additive, the predictors were
generated from a Fourier basis but approximated using a spline basis,
and the nonlinear functions, $g_1(x)$ and $g_2(x)$, were generated
according to sin and cos functions, respectively, but approximated
using a spline basis. In addition, all the various FRAME tuning
parameters were automatically selected using CV, as part of the fitting
process, so the true dimension of the basis functions was not assumed
to be known.

% & & Mean & Last Obs.& FPCA& FPCA-FAR & FRAME \\
%Functional & FP & NA &0.3449& 0.4925& 0.0000 & 0.0000 \\
% & & NA & (0.0160)& (0.0244) & (0.0000)& (0.0000)\\
% & FN & NA & 0.2773& 0.0600& 0.0000 & 0.0000\\
% & & NA & (0.0140)& (0.0126)& (0.0000)& (0.0000) \\
%Scalar & FP & NA & 0.3512& 0.3914& 0.0000 & 0.0000 \\
% & & NA & (0.0135)& (0.0163)& (0.0000)& (0.0000) \\
% & FN & NA & 0.0000 & 0.0000 & 0.0000& 0.0000 \\
% & & NA & (0.0000) & (0.0000)& (0.0000) & (0.0000)\\
% & PE & 2.4348 & 1.1886& 0.9304& 0.0995& 0.0871 \\
% & & (0.0048) & (0.0002)& (0.0031)&(0.0013) & (0.0002)\\
%(FN) rates and their prediction errors for the five comparison
%methods, averaged over the $100$ simulation runs. Standard errors are
%provided in parentheses.}

\begin{figure}

\includegraphics{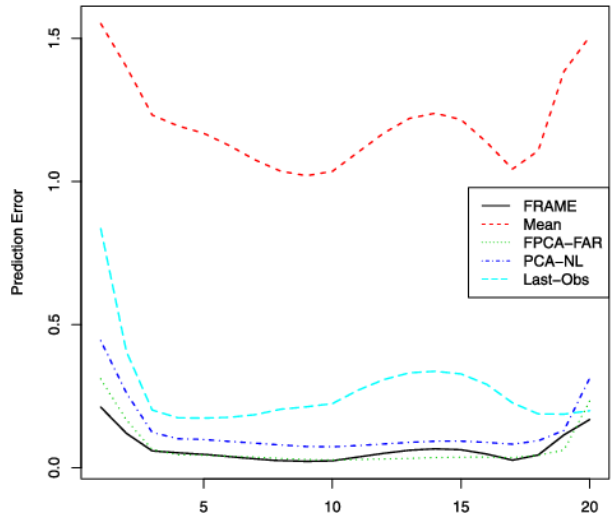}

\caption{Mean prediction errors for five of the comparison methods at
each of the $20$ time points that the response function was observed
over. The Classical and PCA-L curves were not plotted to make the
figure easier to read.}
\label{simplot}
\end{figure}

We generated $100$ different training data sets and fit each of the
seven methods to all $100$ data sets. False negative rates (FN), the
fraction of signal variables incorrectly excluded, and false positive
rates (FP), the fraction of noise variables incorrectly included, were
computed. The prediction error,
$\mathrm{PE} = \frac{1}{20N}\sum_{i=1}^N\sum_{l=1}^{20}(Y_i(s_l) - \widehat
Y_i(s_l))^2$, was also calculated on a large test data set with
$N=1000$ observations. The results, averaged over the $100$
simulations, are displayed in Table~\ref{simres}, with standard
errors shown in parentheses. Since the Last Observation method contains
separate fits for each time point, its FN and FP rates are averaged
over the twenty different fits. Figure~\ref{simplot} plots the
prediction errors over~$s$.

%
%t1 #&#
\begin{table}
\tabcolsep=0pt
\caption{False positive (FP) rates, false negative (FN) rates and
their prediction errors (PE) for the seven comparison methods, averaged
over the $100$ simulation runs. The top rows relate to the functional
predictors, $X_j(t)$, and the lower rows to the scalar predictors,
$Z_k$. Standard errors are provided in parentheses}\label{simres}
\begin{tabular*}{\tablewidth}{@{\extracolsep{\fill}}@{}lcccccccc@{}}
\hline
&& \textbf{Mean} & \textbf{Classical} & \textbf{PCA-L} & \textbf{PCA-NL} & \textbf{Last Obs.} & \textbf{FPCA-FAR} & \textbf{FRAME}\\
\hline
Functional & FP & -- &-- & 0.0600 & 0.4200 & 0.2395 & 0.0375 & 0.0000\\
& & -- & -- & (0.0182) & (0.0333) & (0.0101) & (0.0114) & (0.0000) \\
& FN & -- & -- & 0.4400 & 0.0200 & 0.3002 & 0.1300 & 0.0600 \\
& & -- & -- & (0.0163) & (0.0098) & (0.0102) & (0.0220) & (0.0163)
\\[3pt]
Scalar & FP & -- & -- & 0.0971 & 0.3671 & 0.2419 & 0.0400 & 0.0000 \\
& & -- & -- & (0.0175) & (0.0247) & (0.0089) & (0.0117) & (0.0000) \\
& FN &-- & -- & 0.0000 & 0.0000 & 0.0000 & 0.0000 & 0.0000 \\
& & -- & -- & (0.0000) & (0.0000) & (0.0000) & (0.0000) & (0.0000)
\\[3pt]
& PE & 1.1983 & 0.1108 & 0.1040 & 0.1284 & 0.2727 & 0.0680 & 0.0651 \\
& & (0.0035) & (0.0030) & (0.0029) & (0.0024) & (0.0035) & (0.0019) & (0.0020)\\
\hline
\end{tabular*}
\end{table}

%
%f4 #&#

All methods show significant improvement over the Mean approach,
indicating that the scalar and functional variables have real
predictive ability. FRAME had perfect variable selection results on the
scalar predictors, with false positive and false negative rates both
being zero. All methods had zero false negative rates on the scalar
predictors. However, PCA-NL and Last Observation both had high false
positive rates. FRAME also did a much better job than all its
competitors in identifying the functional predictors. PCA-NL and Last
Observation had high false positive rates for the functional
predictors, and the PCA-L and Last Observation methods had high false
negative rates. In terms of prediction error, FRAME is considerably
superior to all methods except for FPCA-FAR. In comparing FRAME to
FPCA-FAR, we note that while FRAME only results in a small improvement
in terms of prediction error, it does a far better job in selecting the
correct variables.

%s5 #&#
\section{\texorpdfstring{Forecasting demand decay rates.}{Forecasting demand decay rates}}\label{resultssec}

In this section we provide results from applying our FRAME approach to
the HSX data. In doing so, we assume that the revenue curves of any two
movies are independent, given the predictors. This assumption is not
unreasonable because managers use strategic scheduling [\citet
{Einav2010}] to minimize the risk of two movies simultaneously
competing for the same audience. More importantly, the HSX data (i.e.,
our predictors) have incorporated relevant information about the movies
[\citet{foutzjank2010}]. Hence, one might expect much lower
correlations among movies after conditioning on the predictors.

%
%f5 #&#
\begin{figure}

\includegraphics{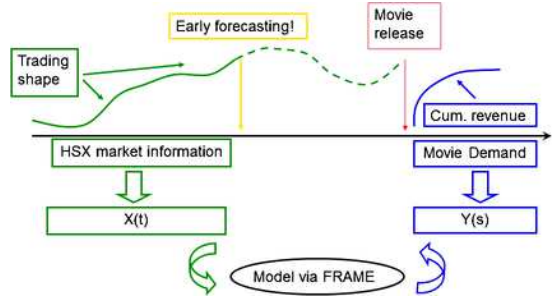}

\caption{Illustration of our model.}
\label{plotillustration}
\end{figure}

Figure~\ref{plotillustration} illustrates the modeling setup.
Recall that for each movie we collect four functional predictors: the
intra-day average price, the number of accounts shorting the stock, the
number of shares sold and the number of shares held short. These curves
capture related yet distinct aspects of
consumer sentiment and word of mouth about a movie.
%number of accounts trading, number of shares traded, number of shares
%covered, dollar volume traded, dollar volume shorted, and dollar
%volume held short.
%The original data contains a number of additional predictors but
%several of them are highly correlated with the intra-day average price
%variable. The best results are obtained using the predictors which are
%least correlated with price.
The four functional predictors (represented using the green curve
before the movie release in Figure~\ref{plotillustration}) are
observed from $52$ up to $10$ weeks prior to the movie's release. We
then use FRAME to form predictions of $Y_i(s)=\log(\mbox{cumulative
revenue for}\break \mbox{movie $i$ at week $s$})$ (blue line after the movie release).

%
%t2 #&#
\begin{table}[b]
\tabcolsep=0pt
\caption{Mean absolute errors (MAEs) on test data for FRAME and six
competing methods averaged over twenty random partitions of the movies}\label{predictions}
\begin{tabular*}{\tablewidth}{@{\extracolsep{\fill}}@{}lccccccc@{}}
\hline
& \textbf{Mean} & \textbf{Classical} & \textbf{PCA-L} & \textbf{PCA-NL} & \textbf{Last Obs.} & \textbf{FPCA-FAR} & \textbf{FRAME}\\
\hline
Week 1 & 2.1898 & 1.5365 & 1.5856 & 1.1793 & 1.1534 & 1.2011 & 1.0952 \\
Week 2 & 2.0490 & 1.4214 & 1.4582 & 1.0951 & 1.0683 & 1.1165 & 1.0116 \\
Week 3 & 1.9057 & 1.3107 & 1.3372 & 1.0157 & 1.0335 & 1.0323 & 0.9482 \\
Week 4 & 1.8335 & 1.2694 & 1.2900 & 0.9915 & 0.9970 & 1.0106 & 0.9364 \\
Week 5 & 1.7907 & 1.2490 & 1.2666 & 0.9815 & 0.9923 & 1.0002 & 0.9305 \\
Week 6 & 1.7610 & 1.2385 & 1.2527 & 0.9785 & 0.9944 & 0.9960 & 0.9324 \\
Week 7 & 1.7418 & 1.2329 & 1.2431 & 0.9759 & 0.9868 & 0.9952 & 0.9371 \\
Week 8 & 1.7294 & 1.2301 & 1.2379 & 0.9749 & 1.0132 & 0.9947 & 0.9397 \\
Week 9 & 1.7199 & 1.2269 & 1.2337 & 0.9759 & 0.9938 & 0.9952 & 0.9432 \\
Week 10 & 1.7144 & 1.2261 & 1.2322 & 0.9772 & 1.0051 & 0.9962 & 0.9460\\
\hline
\end{tabular*}
\end{table}

In Section~\ref{predsec} we test the predictive accuracy of FRAME on
the HSX data in relation to that of several competing methods. Then in
Section~\ref{insightsec} we discuss a graphical approach to obtain
new insight into the relationship between VSMs and movies' success.

%s5.1 #&#
\subsection{\texorpdfstring{Prediction accuracy.}{Prediction accuracy}}\label{predsec}

We compare a number of functional and nonfunctional methods to predict
the %10-week
box office cumulative revenue pattern for our $262$ movies. Table~\ref
{predictions} provides weekly mean absolute errors (MAE) between the
predicted and actual cumulative box office revenue (on the log scale)
for FRAME as well as six comparison methods. Specifically, we randomly
divide the movies into training and test data (180 and 82 movies,
resp.), fit the various methods using the training data and then
compute MAE for week $s$ on the test data:
%
%e14 #&#
\begin{equation}
\label{mae} \mathrm{MAE}(s) =\frac{1}{\llvert  \mathcal T\rrvert  }\sum_{i\in\mathcal T}
\bigl\llvert Y_{i}(s) - \widehat Y_i(s)\bigr\rrvert,
\end{equation}
where $\mathcal T$ represents the test data and $\widehat Y_{i}(s)$ the
prediction for week $s$ using a given method. We repeat this process
over $20$ random partitions of the movies and average the resulting
MAE's. All seven methods are implemented in the same fashion as was
used in the simulation analysis.

A few trends are clear from Table~\ref{predictions}. First, all
methods dominate Mean, indicating that the HSX curves contain useful
predictive information. Second, the errors tend to decline over time,
suggesting that there is more variability in the early weeks, but, to
some extent, this averages out over time. Third, PCA-NL, FPCA-FAR and
Last Observation give similar results and dominate Classical and PCA-L.
Thus, there is clear evidence of a nonlinear relationship. % and that
%using the entire predictor curve provides additional information over
%just adopting the last observed value, so the HSX market is not fully
%efficient.
Finally, FRAME provides superior results in comparison to the other six
approaches for each of the ten weeks. The relative advantage of FRAME
is highest in the first couple of weeks where predictions appear to be
the most difficult. %The advantage is particularly strong in the early
%weeks which appear to be the hardest on which to form accurate
%predictions.
%Figure~\ref{plotcumplot} provides a graphical representation of the
%MAE's from Table~\ref{predictions}.

% \label{plotcumplot}

%
%t3 #&#
\begin{table}
\tabcolsep=0pt
\caption{Average number of times each of the four predictors were
selected for each method}\label{varselect}
\begin{tabular*}{\tablewidth}{@{\extracolsep{\fill}}@{}lcccc@{}}
\hline
& \textbf{Price} & \textbf{Account short} & \textbf{Shares sold} & \textbf{Shares short}\\
\hline
FRAME & 1.00 & 1.00 & 0.00 & 0.00 \\
FPCA-FAR & 1.00 & 0.30 & 0.00 & 0.00 \\
PCA-L & 1.00 & 0.80 & 0.00 & 0.00 \\
PCA-NL & 1.00 & 0.05 & 0.30 & 0.65 \\
Last. Obs. & 1.00 & 0.58 & 0.62 & 1.00\\
\hline
\end{tabular*}
\end{table}

Table~\ref{varselect} records the number of times each of the four
predictors were selected, averaged over the $20$ different training
data sets. The intra-day average price variable appears to be the most
important, with all methods selecting it on every run. FRAME also
selected the variable of accounts trading short but ignored the
remaining two predictors. By comparison, Last Observation chose the
largest models, often including all four predictors. This may have been
to compensate for the fact that the method only observed the final time
point for each curve.

%
%t4 #&#
\begin{table}
\tabcolsep=0pt
\caption{Mean absolute errors on test data using various
characteristics of the movies. Errors are averaged over twenty random
partitions}\label{alternatives}
\begin{tabular*}{\tablewidth}{@{\extracolsep{\fill}}@{}lccccccc@{}}
\hline
& \textbf{Genre} & \textbf{Sequel} & \textbf{Budget} & \textbf{Rating} & \textbf{Run time} & \textbf{Studio} & \textbf{All}\\
\hline
Week 1 & 1.632& 2.136& 1.899& 1.850& 2.209& 2.040& 1.445\\
Week 2 & 1.589& 2.003& 1.762& 1.749& 2.064& 1.915& 1.395\\
Week 3 & 1.510& 1.858& 1.620& 1.634& 1.905& 1.770& 1.312\\
Week 4 & 1.487& 1.792& 1.564& 1.604& 1.829& 1.714& 1.304\\
Week 5 & 1.472& 1.753& 1.535& 1.587& 1.784& 1.685& 1.296\\
Week 6 & 1.463& 1.728& 1.516& 1.578& 1.755& 1.668& 1.291\\
Week 7 & 1.458& 1.713& 1.501& 1.569& 1.735& 1.656& 1.287\\
Week 8 & 1.457& 1.703& 1.492& 1.563& 1.723& 1.648& 1.286\\
Week 9 & 1.457& 1.695& 1.487& 1.561& 1.714& 1.642& 1.287\\
Week 10 & 1.458& 1.691& 1.484& 1.559& 1.709& 1.639& 1.287\\
\hline
\end{tabular*}
\end{table}

To further benchmark FRAME against alternative methods that are
commonly used in the literature on movie demand forecasting [\citet
{SawhneyEliashberg1996}], Table~\ref{alternatives} provides error
rates for seven additional models. For each of these models, we
estimate ten separate weekly linear regressions, one for each of the
ten revenue weeks. We fit each regression to the training data, using
the same $20$ random partitions as in Table~\ref{predictions}, and
report the average MAE's on the test data. The first six models are
based on individual movie features, respectively, genre (e.g., drama or
comedy), sequel (yes/no), production budget (in dollars), MPAA rating,
run time (in minutes) and studios (e.g., Universal or 20th Century
Fox). The seventh model is based on a
combination of all six features. The best individual predictor appears
to be genre, but combining all six predictors gives the best results.
However, the MAE's from the combined model are still significantly
higher than for the best methods in Table~\ref{predictions},
suggesting that the HSX curves provide additional information beyond
that of the movie features.

\subsubsection{\texorpdfstring{Why does FRAME predict so well?}{Why does FRAME predict so well}}

We now offer a closer look into when (and potentially why) the
prediction accuracy of FRAME is superior to that of the alternative
methods in Tables~\ref{predictions} and \ref{alternatives}.
To that end, we investigate the relationship between FRAME's mean
absolute percentage error (MAPE) in cumulative revenues over the first
ten weeks since release %prediction error (i.e. the mean absolute error
%between FRAME's predicted log cumulative revenue and their true values)
and film characteristics, such as budget, genre, MPAA rating, and the
volume and valence of critics' reviews. Similarly, we examine how the
relative performance of FRAME (i.e., the difference between FRAME's
MAPE and the lowest MAPE of either PCA-NL or FPCA-FAR) is associated
with film characteristics.
Tables~\ref{reg1} and \ref{reg2} show the linear regression results.
%models obtained from variable selection via stepwise regression (using
%a combination of forward selection and backward elimination).

%
%t5 #&#
\begin{table}
\tablewidth=270pt
\caption{Linear regression of FRAME's prediction error on film characteristics}\label{reg1}
\begin{tabular*}{\tablewidth}{@{\extracolsep{\fill}}@{}ld{2.3}cd{2.3}c@{}}
\hline
\textbf{Name} & \multicolumn{1}{c}{\textbf{Coefficient}} & \multicolumn{1}{c}{\textbf{Std err.}} & \multicolumn{1}{c}{$\bolds{t}$} & \multicolumn{1}{c@{}}{$\bolds{p}$\textbf{-value}}\\
\hline
Intercept & 0.098 & 0.068 & 1.439 & 0.151 \\
Sequel &-0.033 & 0.014 & -2.314 & 0.022 \\
Budget & 0.000 & 0.000 & 0.587 & 0.558 \\
Action & -0.015 & 0.050 & -0.296 & 0.768 \\
Animated & 0.016 & 0.054 & 0.306 & 0.760 \\
Comedy & -0.009 & 0.050 & -0.185 & 0.853 \\
Drama & -0.011 & 0.050 & -0.216 & 0.829 \\
Horror & 0.004 & 0.050 & 0.086 & 0.931 \\
Other genres & 0.066 & 0.060 & 1.098 & 0.273 \\
Rating below R & -0.026 & 0.011 & -2.417 & 0.016 \\
Runtime & 0.001 & 0.000 & 2.516 & 0.013 \\
Major studio & -0.039 & 0.010 & -3.744 & 0.000 \\
Oscar & 0.030 & 0.028 & 1.062 & 0.289 \\
Critics volume & -0.001 & 0.000 & -7.600 & 0.000 \\
Critics valence & 0.006 & 0.005 & 1.322 & 0.188 \\
Consumer WOM volume & 0.000 & 0.000 & 1.943 & 0.053 \\
Consumer WOM valence & 0.004 & 0.006 & 0.654 & 0.514\\
\hline
\end{tabular*}
\end{table}

%
%t6 #&#
\begin{table}
\tabcolsep=0pt
\tablewidth=270pt
\caption{Linear regression of the difference between FRAME's
prediction error and the lowest error of either PCA-NL or FPCA-FAR on
film characteristics}\label{reg2}
\begin{tabular*}{\tablewidth}{@{\extracolsep{\fill}}@{}ld{2.3}cd{2.3}c@{}}
\hline
\textbf{Name} & \multicolumn{1}{c}{\textbf{Coefficient}} & \multicolumn{1}{c}{\textbf{Std err.}} & \multicolumn{1}{c}{$\bolds{t}$} & \multicolumn{1}{c@{}}{$\bolds{p}$\textbf{-value}}\\
\hline
Intercept & 0.011 & 0.024 & 0.465 & 0.642 \\
Sequel & 0.000 & 0.005 & 0.036 & 0.971 \\
Budget & 0.000 & 0.000 & -0.307 & 0.759 \\
Action & -0.013 & 0.018 & -0.746 & 0.456 \\
Animated & -0.019 & 0.019 & -1.023 & 0.308 \\
Comedy & -0.018 & 0.017 & -1.010 & 0.314 \\
Drama & -0.023 & 0.017 & -1.326 & 0.186 \\
Horror & -0.012 & 0.018 & -0.685 & 0.494 \\
Other genres & 0.015 & 0.021 & 0.721 & 0.471 \\
Rating below R & -0.001 & 0.004 & -0.302 & 0.763 \\
Runtime & 0.000 & 0.000 & -1.101 & 0.272 \\
Major studio & -0.006 & 0.004 & -1.731 & 0.085 \\
Oscar & 0.017 & 0.010 & 1.764 & 0.079 \\
Critics volume & 0.000 & 0.000 & 3.198 & 0.002 \\
Critics valence & -0.001 & 0.002 & -0.533 & 0.595 \\
Consumer WOM volume & -0.000 & 0.000 & -3.901 & 0.000 \\
Consumer WOM valence & 0.004 & 0.002 & 1.936 & 0.054\\
\hline
\end{tabular*}
\end{table}

Table~\ref{reg1} shows that FRAME performs well (i.e., has a low
prediction error) for movies that are sequels, rated below $R$,
have a shorter runtime, are released by a major studio such as
Paramount, Warner Brothers, Universal or Twentieth Century Fox, and
reviewed by a larger number of critics.
Intuitively, these results suggest that FRAME performs especially well
for movies that enjoy a greater capability for creating pre-release buzz.
For instance, sequels build upon the success of their predecessors;
films released by major studios benefit from significant advertising
and publicity before opening;
those with lower MPAA ratings, for example, G~and~PG, appeal to wider
audiences; and greater attention from the critics, due to, for
instance, a film's quality or controversies, could further fuel the
public's fascination.
Such firm- or consumer-generated buzz provides rich information to the
HSX traders, who rapidly integrate the information into the stock trading.
FRAME seems to be capable of capturing the dynamics of such buzz and
translating it into accurate predictions.

\begin{figure}

\includegraphics{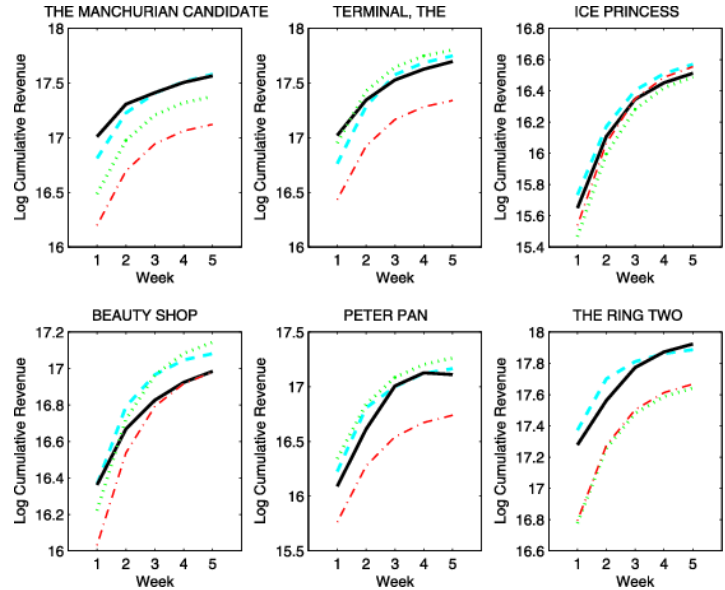}

\caption{Top 6 movies with the smallest FRAME prediction error: the
solid lines correspond to FRAME's prediction; the dashed lines show the
corresponding true values. The two closest competitors are given by the
dotted lines (PCA-NL) and the dash-dotted lines (FPCA-FAR),
respectively.} \label{plotsmallesterrors}
\end{figure}

Figure~\ref{plotsmallesterrors} shows the six movies for which FRAME
predicts the best in terms of MAPE. Two-thirds of these six movies were
released by major studios with the
exception of \textit{THE RING TWO} and \textit{THE TERMINAL}. Moreover, all
of them were rated below R except for \textit{THE MANCHURIAN CANDIDATE}.
And all attracted more than a hundred critics' reviews.
A third of them are sequels, specifically \textit{PETER PAN} and \textit{THE
RING TWO}, as compared to 11\% in the sample. Moreover, sequels are not
far down the list. For example, FRAME also provides excellent
predictions for
sequels like \textit{MISS CONGENIALITY 2} and \textit{OCEAN'S TWELVE}. By
contrast, FRAME predicts the least accurately for the following movies:
\textit{KAENA}: \textit{THE PROPHECY}, \textit{THE INTENDED} and \textit{EULOGY}. None of
these movies was a sequel or produced by a major studio. Only \textit{KAENA}: \textit{THE PROPHECY} had a below-R rating; and the volumes of critics'
reviews for all three movies were below 35.

% \caption{Dependence plots for different input shapes.}
% \label{plotpartialdepplot}

%
%f6 #&#

It is possible that movies with some of the above identified
character\-istics---sequels, low MPAA ratings, major studio releases and
more critics' reviews---are easier to predict in general by any
method, not only by FRAME.
Indeed, Table~\ref{reg2} shows that FRAME does not have a
statistically significant advantage (despite directionally so) over
PCA-NL or FPCA-FAR in
predicting demand for films of the above characteristics. Nonetheless,
FRAME continues to outperform the alternative methods for films
generating more viewer ratings online, suggesting its distinct
ability to incorporate information potentially not captured by
alternative methods, such as potential viewers' interest that is not
widely available ten weeks prior to a film's release.

%s5.2 #&#
\subsection{\texorpdfstring{Model insight.}{Model insight}}\label{insightsec}

%Observation Non-Linear (red short dash), FPCA Non-Linear (blue dash
%dot) and true revenues (cyan long dash) for a sample of six movies.
%The top row corresponds to the three moves FRAME predicted most
%accurately and the bottom row to those it was least accurate for.}
% \label{plotindivcum}

The previous section has shown that using a fully functional regression
method such as FRAME can be beneficial for forecasting demand decay
patterns. However, while nonlinear functional regression methods can
result in good predictions, one downside is that because both
model-input (HSX trading paths) as well as model-output (cumulative box
office demand) arrive in the form of functions, it is hard to
understand the relationship between the response and the predictors.
%[GARETH, YINGYIING: MAYBE ELABORATE IN ONE OR TWO MORE SENTENCES WHY
%FOR THE FRAME MODEL IT IS IMPOSSIBLE TO "SEE" THE RELATIONSHIP BETWEEN
%INPUT AND OUTPUT CURVES?]
%In fact, since the model operates
%on principal components underlying the input- and output-shapes, it
%is almost impossible to ``see'' the relationship between the two.

A useful graphical method to address this shortcoming is to visualize
the relationship by generating candidate predictor curves, using the
fitted FRAME model to predict corresponding responses and then plotting
$X(t)$ and $Y(s)$ together.
%To overcome this drawback, we develop new visualizations that transform
%principal components back to their original domain and, as a result,
%allow for insightful conclusions.
The idea is similar to the ``partial dependence plots'' described in
\citet{HastieTibFriedman2001}; however, in contrast to their
approach, our plots take into account the joint effect of all
predictors (and are hence not ``partial''); we thus call our graphs
``dependence plots.''

%{flowchart.eps}}
% \caption{Illustration of reverse-engineering output shapes
% from an idealized input shape.}
% \label{plotflowchart}
%
%Recall that our model links input shapes (HSX trading paths) to
%output shapes (demand decay rates). We operationalize it by first
%decomposing both input and output into their most representative
%principal components (PCs), and then modeling the relationship
%between input and output PCs in a multivariate fashion. The top
%panel in Figure \ref{plotflowchart} illustrates this process.

%
%f7 #&#
\begin{figure}

\includegraphics{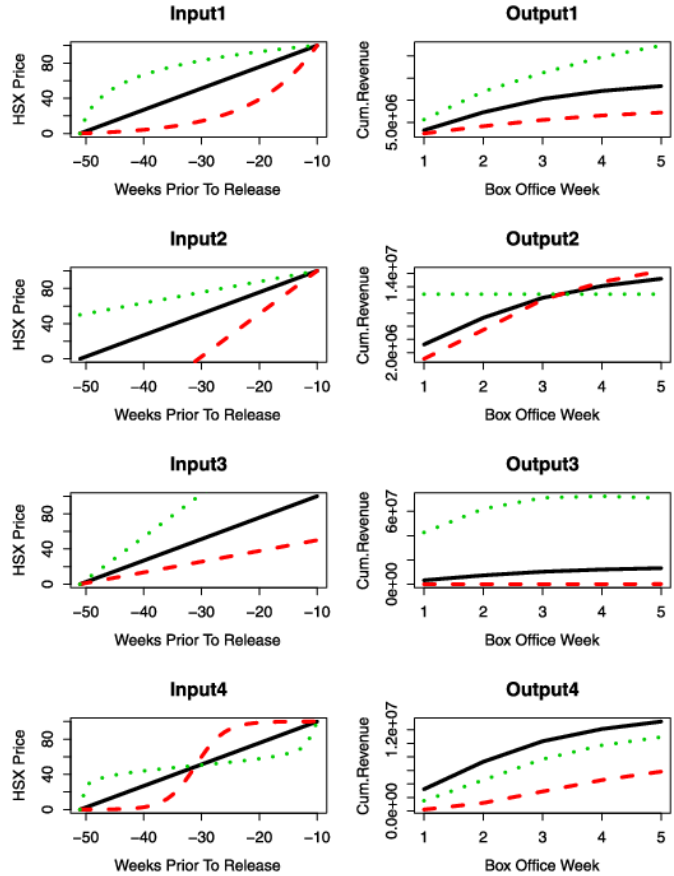}

\caption{Dependence plots for different input shapes. The left panels
contain various idealized input curves of HSX prices over time. Each
figure plots three possible shapes for the observed HSX trading history
of a movie. The right panels plot the corresponding predicted
cumulative revenues using FRAME. For example, in the top row we observe
that an HSX trading curve which increases rapidly and then levels off
(dotted line) corresponds to a higher predicted revenue than either a
linear pattern (solid line) or slow start with a large increase at the
end (dashed line).}
\label{plotpartialdepplot}
\end{figure}

Figure~\ref{plotpartialdepplot} displays several possible
dependence plots with idealized input curves in the left panel and
corresponding output curves from FRAME in the right panel. Note that
since in our empirical analysis the intra-day average
price was by far the most important predictor, we use that variable as
$X(t)$ and fit FRAME with this single functional predictor. We study a
total of four different scenarios. The top row corresponds to a
situation where all input curves start and end at the same values (0
and 100, resp.); their only difference is how they get from the
start to the end: the middle curve (solid line) grows at a linear rate;
the upper and lower curves (dotted and dashed lines) grow at
logarithmic and exponential rates, respectively. In that sense, the
three curves represent movies whose HSX prices either grow at a
constant (linear) rate, or grow fast early but then slow down
(logarithmic) or grow slowly early only to increase toward release
(exponential).
%Movies that grow fast early may be sequels who often enjoy early
%awareness from their predecessors; on the other hand, slow early
%growth with faster increases towards the release date may be a
%trademark of so-called ``sleeper'' movies.
%All three curves end at exactly the same HSX market value, so any
%difference in estimated box office revenue is only due to their
%difference in shapes.

The top right panel shows the result: the logarithmic HSX price curve
(dotted line) results in the largest cumulative revenue. In particular,
its cumulative revenue is larger compared to the linear price curve
(solid line), and both logarithmic and linear price curves beat the
cumulative revenue generated by the exponential price curve (dashed
line). In fact, the logarithmic price curve results in cumulative
revenue that continues to grow significantly, especially in later
weeks. This is in contrast to the cumulative revenue generated by the
exponential price curve which becomes almost constant after week two or three.

% at any point in time and especially in later weeks (weeks 3 to 5).
%similar decay of box office revenue, and both significantly
%out-perform the exponential curve (red dashed line). In fact, while
%both linear and logarithmic HSX price growth result in very high box
%office revenues during the first few weeks, exponential price growth
%leads to very low revenues (which stay low). Comparing linear and
%logarithmic price growth with one another, we notice that the
%logarithmic shape results in a slight revenue advantage during the
%first few weeks.

What do these findings imply? Recall that all three HSX price curves
start and end at the same value (0 and 100, resp.), so all
observed differences are only with respect to their shape.
This suggests that shapes matter enormously in VSMs. It also suggests
that more buzz early on (i.e., the logarithmic shape) has much more
impact on the overall revenue compared to a last moment hype closer to
release time (i.e., the exponential shape).

The next two rows of Figure~\ref{plotpartialdepplot} show
additional shape scenarios with both rows displaying input curves with
a common linear shape. In the second row the curves are converging
toward a common HSX value, while the input curves in the third row are
diverging.
%While the diverging input curves result in very different cumulative
%revenue curves (especially with respect to their value in week 5), the
%cumulative revenue curves based on the converging input curves share
%more similarity, especially with respect to their week 5 value.
The case of diverging curves suggests that the larger the most recent
HSX value, the larger is the corresponding cumulative box office
revenue. The converging case emphasizes the effect of recency of
information: like in panel~1, all HSX price curves end at the same
value; however, unlike in panel~1, they all have the same shape. We can
see that the corresponding cumulative box office revenue also almost
converges in week 5. This suggests that the difference in shape (e.g.,
linear vs. logarithmic vs. exponential) carries important information
about the \textit{change} in the dynamics of word of mouth or
consumer-generated buzz which translates into significant revenue differences.

The last row in Figure~\ref{plotpartialdepplot} shows yet another
scenario of HSX price curves: an S-shape (dashed line) and an inverse-S
shape (dotted line). Notice that the inverse-S shape features spurts of
extreme growth both at the very beginning and at the very end, almost
like a combination of logarithmic and exponential growth from panel~1.
However, while the spurts resemble the logarithmic and exponential
shapes, their overall magnitude is smaller compared to that in panel~1.
As a result, the cumulative revenue is smaller compared to that of the
linear growth. This suggests that while the dynamics of HSX price
curves matter, their magnitude and timing matters even more, as the
linear HSX price curve features a much more steady and sustained
overall change in HSX prices compared to the inverse-S shape (which is
constant most of the time with two small spurts at the beginning and
the end). More evidence for this can be seen in the S-shaped HSX price
curve (dashed line): while it does feature some change, most of the
change happens in the middle of the curve which leads to the lowest of
the three cumulative revenue curves.
%This again suggests that while dynamics matter, the timing of the
%dynamics in the S-shaped curve (which features all activity in the
%middle) does not lead to significant growth in revenue.

%the S shape features a very prominent period of strong and steady
%growth, one that is similar to (but exceeds in magnitude) the
%logarithmic shape from the first scenario. The result supports our
%previous findings: while the almost-constant growth of the inverse-S
%shape produces very low box office revenues, the strong dynamics of
%the S-shape lead to revenues that exceed those of the logarithmic
%shape in the first scenario.

%All-in-all, our results suggest that, in addition to the magnitude of
%the HSX price curve (which captures level-differences in perception
%about a movie), its shape is capable of capturing information about
%buzz. They also suggest that the timing of the buzz is an important
%predictor of box office success.

%s6 #&#
\section{\texorpdfstring{Conclusion.}{Conclusion}}\label{conclusionsec}
This paper makes three significant contributions. First, we develop a
new nonlinear regression approach, FRAME, which is capable of forming
predictions on a functional response given multiple functional
predictors and simultaneously conducting variable selection. Our
results on both the HSX and simulated data demonstrate that FRAME is
capable of providing a considerable improvement in prediction and
variable selection accuracy relative to a host of competing methods.
Second, we introduce a new and promising data source to the statistics
community. Online virtual stock markets (VSMs) are market-driven
mechanisms to capture opinions and valuations of large crowds in a
single number. Our work shows that the information captured in VSMs is
rich but requires appropriate and creative statistical methods to
extract all available knowledge [\citet{jankshmu2006}].
%we illustrate the power of
%functional modeling for analyzing VSMs by developing an innovative
%model that uses shapes both as input and output values and links
%them using non-linear regression methods. This approach improves
%predictive accuracy relative to a series of competitor models.
Finally, we make our approach practical for inference purposes by
developing dependence plots to illustrate the relationship between
input and output curves.

FRAME overcomes some of the technical difficulties encountered in other
functional models.
For instance, FRAME does not require the calculation of eigenfunctions,
as is the
case with our benchmark method, FPCA, in, for example, Tables~\ref{simres} or \ref{varselect}.
In FPCA, we first compute
the principal components of the response curves, and then apply
standard modeling techniques to the principal component scores.
However, since the response curves are observed with random error, so
are the corresponding eigenfunctions. While approaches for removing
this random variation from the eigenfunctions exist [\citet{Yao2005}],
FRAME does not rely on a principal component decomposition and thus
does not encounter this type of challenge.

Our results have important implications for managerial practice.
Equipped with the early forecasts of demand decay patterns, studio
executives can make educated decisions regarding weekly advertising
allocations (both before and after the opening weekend), selection
of the optimal release date to minimize competition with films from
other studios and cannibalization of films from the same studio
[\citet{Einav2007}], and negotiation of the weekly revenue sharing
percentages with the theater owners. Studios may be able to better
manage distributional intensity and consumer word of mouth. For
instance, for a movie predicted to have a strong opening weekend but
fast decay afterward, the studio may consider nationwide release,
as opposed to limited or platform release strategies (i.e., from
initial limited release
to nationwide release later on), at the same time
strategically managing potentially negative word of mouth. The
predicted demand decay of a film will also shed crucial light on a
studio's sequential distributional strategies. For example, a studio
may consider delaying (or shortening) a movie's video release or
international release timing if the movie is predicted to have
longevity (or faster decay) in theaters. Given that many academics
have called for serious research on the optimal release timing in
the subsequent distributional channels, such as home videos and
international theatrical markets
[\citet{EliashbergElberseLeenders2006}], and that these channels
represent five times more revenues than the domestic theatrical box
office [MPAA (2007)], our results bear further crucial
implications to the profitability of the motion picture industry.

A potential limitation of our approach is that it may only add value in
inefficient markets where valuable information, above and beyond the
information contained in the final
trading price, is captured by the shape of the trading histories, such
as prices, accounts and shares. However, as
outlined earlier, previous research suggests that VSMs are not fully
efficient. Furthermore, the strong predictive accuracy of our
functional approach provides further empirical validation for this
finding. In addition, the FRAME methodology is applicable
beyond just VSM data. In general, it can be used on any
regression problem involving functional predictors and responses.

%There are several possible future extensions of this work. In terms of
%the data,
We believe there are many other interesting
applications of VSM's to different domains, such as music, TV shows
and video games which all share similar characteristics to movies, such
as frequent introductions of new, unique and experiential products,
pop culture appeal and strong influence of hype on demand. Such
research would be made possible by the increasing availability of
data from VSMs for, for example, books (MediaPredict), music (HSX), TV
shows (Inkling) and video games (SimExchange).

\begin{appendix}\label{app}
%s7 #&#
\section*{\texorpdfstring{Appendix: Algorithm details}{Appendix: Algorithm details}}

%In the initialization step (Step 0.) of this algorithm some of the $
%corresponding predictors do not appear related to the response.
%However, the initialization assumes a linear model. It is conceivable
%that a response that appears unimportant using a linear model will
%become statistically significant using a non-linear model. Hence, if $
%equal to the loading vector of the first principal component of $
%variability in $X_{ij}(t)$ so is the most natural unsupervised
%projection and allows for potential non-linear relationships to be
%detected in Step 2.

For a general penalty function, $\rho(t)$, we use the local linear
approximation method proposed in \citet{ZL08} to solve (\ref
{nonlinfar2}). The penalty function can be approximated as $\rho(\llVert
\bff\rrVert  ) \approx
\rho'(\llVert  \bff^*\rrVert  )\llVert  \bff\rrVert  +C$, where $\bff^*$ is some vector that is
close to $\bff$ and $C$ is a constant. Hence, the only required change
to the FRAME algorithm for optimizing over general penalty
functions is to replace $\lambda$ by $\lambda^*=\lambda\rho'(\llVert  \hf
_j\rrVert  )$ in the calculation of $c_j$ in 2, and replace $\lambda$ by
$\lambda^*=\lambda\rho'(\llVert  \hat\bphi_k\rrVert  )$ in the calculation
of $c_k$\vadjust{\goodbreak} in 5,
% by
%$$c_j =
% \Big(1-\lambda\rho'(\left\Vert \hf_j\right\Vert )/\left\Vert \hat H_jS^H_j\bR_j\right\Vert \Big)_+,$$
where $\hf_j$ and $\hat\bphi_k$ represent the most recent
estimates for $\bff_j$ and $\bphi_k$. The initial estimates of $\hf
_j$ and $\bphi_k$ can be obtained by using the Lasso penalty. This
simple approximation allows the FRAME algorithm to be easily applied to
a wide range of penalty functions.

To implement the second step of the FRAME algorithm, we minimize (\ref
{etaopt}) with respect to the $\bfeta_j$'s. Directly minimizing (\ref
{etaopt}) is difficult due to the nonlinearity of the functions
$g_j(x)\approx\bh(x)^T\bxi_j$. To overcome this difficulty, we
observe that, with the estimates $\hbxi_j$ and $\hat\balpha_k$ from
Algorithm~\ref{alg1} and the current value, $\bfeta_{j,\mathrm{old}}$, of $\bfeta_j$,
the first order approximation of $g(\tilde\btheta{}_{ijl}^T\bfeta
_j)\approx\bh(\tilde\btheta{}_{ijl}^T\bfeta_j)^T\hbxi_j$ is
\[
\bh\bigl(\tilde\btheta{}_{ijl}^T\bfeta_j
\bigr)^T\hbxi_j\approx\bh\bigl(\tilde
\btheta{}_{ijl}^T\bfeta_{j,\mathrm{old}}\bigr)^T
\hbxi_j + \bh'\bigl(\tilde\btheta{}_{ijl}^T
\bfeta_{j,\mathrm{old}}\bigr)^T\hbxi_j \cdot\tilde
\btheta{}_{ijl}^T(\bfeta _j-\bfeta_{j,\mathrm{old}}).
\]
Thus, we can approximate (\ref{etaopt}) by
%
%e15 #&#
\begin{equation}
\label{etaoptapprox} \sum_{i=1}^n \sum
_{l=1}^{n_i} \Biggl(R_{il} - \sum
_{j=1}^p\bh'\bigl(\tilde
\btheta{}_{ijl}^T\bfeta_{j,\mathrm{old}}\bigr)^T
\hbxi_j \cdot\tilde\btheta{}_{ijl}^T(
\bfeta_j-\bfeta_{j,\mathrm{old}}) \Biggr)^2,
\end{equation}
where $R_{il} = Y_{il} - \sum_{j=1}^p \bh(\tilde\btheta{}_{ijl}^T\bfeta_{j,\mathrm{old}})^T\hbxi_j-\sum_{k=1}^q \bomega_{ikl}^T\hat
\balpha_k $. The above approximation (\ref{etaoptapprox}) is a
quadratic function of $\bfeta_j$ and can be minimized easily. Hence,
the new value of $\bfeta_j$ is updated as the minimizer of (\ref
{etaoptapprox}). We also note that if the estimate~$\hbxi_j$ from
Algorithm~\ref{alg1} is $\bzero$, then the corresponding value of $\bfeta_j$
will not be updated.
\end{appendix}

%
%
%In the initialization step (Step 0.) of this algorithm some of the $
%corresponding predictors do not appear related to the response.
%However, the initialization assumes a linear model. It is conceivable
%that a response that appears unimportant using a linear model will
%become statistically significant using a non-linear model. Hence, if $
%equal to the loading vector of the first principal component of $
%variability in $X_{ij}(t)$ so is the most natural unsupervised
%projection and allows for potential non-linear relationships to be
%detected in Step 2.
%
%To implement Step 3. of the FAR algorithm, we minimize (\ref{etaopt})
%with respect to the $\bfeta_j$'s. Directly minimizing (\ref{etaopt})
%is difficult due to the nonlinearity of the functions $g_j(t)\approx
%estimate $\hbxi_j$ from Step 2. and the current value $\bfeta_{j,\mathrm{old}}$
%of $\bfeta_j$, the first order approximation of $g(\btheta{}_{ij}\t
%Thus we can approximate (\ref{etaopt}) as
%where $R_i = Y_i - \sum_{j=1}^p \bh(\btheta{}_{ij}\t\bfeta_{j,old})\t
%2. of the algorithm in the current iteration. The above approximation (
%minimized easily. Hence the new value of $\bfeta_j$ is updated as the
%minimizer of (\ref{etaoptapprox}). We also note that if the estimate $


%suskaldyti doi

% imsref loaded by linak, 2014-09-25 11:15:39

\printaddresses
\end{document}